\documentclass[reprint,
nofootinbib,
amsmath,amssymb,
aps,
pra
]{revtex4-1}

\usepackage{graphicx}  
\usepackage{dcolumn}   
\usepackage{bm}        
\usepackage{amssymb}   
\usepackage{subcaption}
\usepackage{amsmath}
\usepackage{color}
\usepackage{soul}
\usepackage[hypertexnames=false]{hyperref}
\usepackage{footnote}

\newcommand{\beq}{\begin{equation}}
\newcommand{\eeq}{\end{equation}}
\newcommand{\bea}{\begin{eqnarray}}
\newcommand{\eea}{\end{eqnarray}}

\newcommand{\nn}{\nonumber\\}

\newcommand\fig[1]     {Fig.\,{\ref{#1}}}
\newcommand\sect[1]    {Sec.\,{\ref{#1}}}

\newcommand\app[1]     {Appendix~\ref{#1}}

\newcommand\bol[1] {{\bf{#1}}}
\newcommand\bsy[1] {{\boldsymbol{#1}}}

\renewcommand{\Im}{\,\textrm{Im}\,}
\renewcommand{\Re}{\,\textrm{Re}\,}

\newcommand{\ket}[1]{\left|{#1}\right\rangle}

\newcommand \be {\begin{equation}}
\newcommand \ee {\end{equation}}
\newcommand \bed {\begin{displaymath}}
\newcommand \eed {\end{displaymath}}

\newcommand{\bit}{\begin{itemize}}
	\newcommand{\eit}{\end{itemize}}

\def\eq#1{(\ref{#1})}
\def\s0#1#2{\mbox{\small{$ \frac{#1}{#2} $}}}
\def\0#1#2{\frac{#1}{#2}}

\def\eq#1{(\ref{#1})}
\DeclareMathAlphabet\mathbfcal{OMS}{cmsy}{b}{n}
\begin{document}

\title{Quasiparticles in an interacting system of charge and monochromatic field}

\author{P. Mati}

\email{Peter.Mati@eli-alps.hu}
\email{matipeti@gmail.com}
\affiliation{ELI-ALPS, ELI-Hu NKft, Dugonics t\'er 13, Szeged 6720, Hungary}

\date{\today}

\begin{abstract}
The spectrum of an exactly solvable non-relativistic system of a charged particle interacting with a quantized electromagnetic mode is studied with various polarizations. Quasiparticle dispersion relations can be derived from the diagonalized Hamiltonian which, in the case of a linearly polarized field, indicates a bulk plasmon excitation, whereas in the elliptically and circularly polarized cases the dispersions exhibit a global minimum and show a singular behavior as the wavenumber tends to zero. These new type of dispersion relations lead to modified plasma frequencies and reflectivities, as well as to negative group velocities. It is shown that the zero-point energy of the system implies a repulsive force between two parallel plates, which vanishes when the charge is set to zero.
\end{abstract}

\maketitle

\section{Introduction}
\vskip -.08cm
Achieving exact solutions of a problem related to an interacting physical system is rather rare (e.g., \cite{Varro1,Varro2,Varro3,Varro4,Exactsol,JC,Dicke}). These systems are usually simplifications of more complicated situations and are analyzed in order to capture the entire physics and mathematics of the problem under consideration. The physical relevance of the simplified models are limited; however, sometimes it is highly valuable to trade the range of answerable questions for a non-approximated insight on specific properties of the given problem.
In this paper, a non-relativistic system consisting of a charged particle and a quantized mode of radiation field is studied. Considering this interaction with only one mode could seem to be an oversimplification; however, there are physical situations where indeed an approximation of the radiation field with one monochromatic component is well justified, e.g., in lasers. In fact, the same approximations are used in the two fundamental models of quantum optics: the Jaynes-Cummings and the Dicke models \cite{JC,Dicke}. In the work of Bergou and Varr\'o \cite{Varro1}, the corresponding Hamiltonian is exactly diagonalized which provides the non-perturbative description of its spectrum: It describes a free charged particle and a quasi excitation. The relativistic treatment of the same problem can be found in \cite{Varro2}. Varr\'o also studied the entangled photon-electron states of this system in \cite{Varro3,Varro4}.\\
In the present work it is shown that the dispersion relation of the quasi excitations qualitatively depends on the polarization of the field. For linear polarization (LP) it is described by a bulk plasmon dispersion whereas for circular polarization (CP) and elliptic polarization (EP) the dispersion relation has a global minimum but is singular for vanishing wavenumber. The analysis of the dispersion relations shows that the plasma frequency can be different for various polarizations and the group velocities for non-LP cases can acquire negative values. Associated to the minimum of the dispersion relations, the zero-point energy of the system can be found, which implies a repulsive force between two parallel plates.

\section{The model}

The Hamiltonian of the simple interacting system of the charged particle with the quantized mode has the following form 
\bea\label{startHam}
H&=&\frac{1}{2 m}\left(\bol{p}-\frac{e}{c} \bol{A}\right)^2+\hbar\omega\left(\frac{1}{2}+a^\dagger a\right),
\eea 
where $\bol{p}$, $m$, and $e$ are the momentum, the mass, and the charge of the particle; $\omega$ is the angular frequency of the electromagnetic (EM) mode; $c$ is the speed of light; and $\bol{A}$ is the vector potential. Likewise in \cite{Varro1}, the polarization of the electromagnetic field is treated as a parameter of the Hamiltonian and thus two essentially different cases can be distinguished:
\bea\label{vpcl}
\bol{A}_c=\alpha(\bsy{\varepsilon} a+\bsy\varepsilon^{*} a^\dagger),\qquad
\bol{A}_l=\alpha \bsy\epsilon\left(a+a^\dagger\right),
\eea
where $\bol{A}_c$ and $\bol{A}_l$ correspond to the CP and LP cases, respectively. The parameter $\alpha =c \sqrt{2 \pi  \hbar / V \omega }$, with quantization volume $V$ and Planck constant $\hbar$. In the case of CP, the polarization vectors are complex valued with the criteria $\bsy\varepsilon\bsy\varepsilon=\bsy\varepsilon^{*}\bsy\varepsilon^{*}=0$ and $\bsy\varepsilon\bsy\varepsilon^{*}=1$, whereas $\bsy\epsilon$ is a real-valued unit vector for the LP field.\\
In addition to these two, a third case can be introduced corresponding to the EP field
\bea\label{elli}
\bol{A}_e=\alpha(\mathbfcal{E} \,a+\mathbfcal{E}^{*}\,a^\dagger),
\eea
where $\mathbfcal{E}$ must be parametrized so that in certain limits it gives the polarization vectors of the CP and LP cases, i.e., $\mathbfcal{E}(\xi\to1)=\bsy\varepsilon$ and $\mathbfcal{E}(\xi\to0)=\bsy\epsilon$ with some parameter $\xi\in[0,1]$. In particular, the parametrization, $\mathbfcal{E}=\sqrt{1/(1+\xi^2)}(\Re\bol{u}+i\xi\Im\bol{u})$ with $\bol{u}=(1,i)$  gives the right limits and its norm $\mathbfcal{E}\mathbfcal{E}^*=1$, independently of $\xi$. In the following, indices like those in \eqref{vpcl} and \eqref{elli} are suppressed for the sake of clarity, and the EP case is understood unless stated otherwise. Even though in the definition of the vector potential in \eqref{vpcl} and \eqref{elli} a dipole approximation is used, all the results about the dispersion relations can be generalized to plane waves for specific circumstances, as shown in \app{pwgen}. However, it is only important when the properties that are related to the wave propagation are considered.\par
A displacement and a Bogoliubov transformation of the creation and annihilation operators are used for the diagonalization of the Hamiltonian in \eqref{startHam} in order to eliminate the linear and the quadratic terms in $a$ and $a^\dagger$ \cite{Varro1,Varro2}. In the following, only the final result, i.e. the diagonalized Hamiltonian, is presented. The details of the computation can be found in \app{app1}. The EP case is the most general and is the focus of all the results presented but both the CP and the LP results can be obtained by taking the appropriate limits in $\xi$. The Hamiltonian after the transformation reads
\bea\label{hamil}
\mathcal{H}&=&\frac{{\bf p}^2}{2m}+\hbar \Omega\left(b^\dagger b+\frac{1}{2}-\sigma^\dagger \sigma\right),
\eea
 where $\Omega$ plays the key role in the differences between the spectra for the differently polarized modes:
\bea\label{omegae}
\Omega=\sqrt{\omega ^2+\omega_p^2 \left(1+\frac{\xi^2}{\left(\xi^2+1\right)^2}\frac{\omega_p^2}{\omega^2}\right)},
\eea
where $\omega_p^2=4 \pi e^2/ m V $ was introduced as "plasma frequency" \cite{Varro1,Multiphoton}. The plasma oscillation usually is considered a collective phenomenon, where a number of $N$ charged particles define the given plasma frequency with $\omega_p^2=4 \pi e^2 n_e/ m $, where $n_e=N/V$ is the electron density \cite{plasma}. However, in the present case $N=1$.
The operators, $b$ and $b^\dagger$, are obtained through a Bogoliubov transformation ($C_\theta=\exp\{\frac{1}{2}\theta(a^\dagger a^\dagger + a a)\}$), which is necessary in order to eliminate the quadratic terms in $a$ and $a^\dagger$:
\bea\label{bogo}
b&=& C_\theta^{-1} b C_\theta =\cosh {\theta} a + \sinh{\theta} a^\dagger,\nn
b^\dagger&=& C_\theta^{-1} b^\dagger C_\theta= \cosh{\theta} a^\dagger + \sinh {\theta} a.
\eea
The argument of the hyperbolic functions is defined through the relation
\bea\label{theta}
\tanh 2 \theta = \frac{e^2 \alpha^2}{ m c^2} \frac{1}{\hbar\omega + \frac{e^2 \alpha^2}{mc^2}}\frac{1-\xi^2}{1+\xi^2}.
\eea 
The transformation in \eqref{bogo} for the special case of the CP ($\xi\to1$) coincides with the identity transformation. The transformation $C_\theta$ is the squeeze operator that generates a squeezed state from the vacuum (in the original Fock space) \cite{Bogol} which will be the vacuum state in the transformed Fock space.
There is an additional shift to the Bogoliubov-transformed number operator ($b^\dagger b=\hat n_b$) in \eqref{hamil}, which consists of the displacement transformation parameter $\left( D_{\sigma} ^{-1}b^{(\dagger)}D_{\sigma}=b^{(\dagger)} + \sigma^{(\dagger)} \right)$ that performs a shift on the raising and lowering operators:
\bea
\!\!\!\!\!\!\! \sigma= \frac{\cosh 2\theta}{\hbar\omega + \frac{e^2 \alpha^2}{mc^2}}\frac{\alpha e}{m c} \frac{{\bol{p}}}{\sqrt{(1+\xi^2)}}\left[e^{-\theta}\Re\bol{u}-e^{\theta}i \xi \Im\bol{u}\right].
\eea
The operator $D_\sigma=\exp(\sigma b^\dagger-\sigma^\dagger b)$ generates a coherent state from the vacuum $\ket{0}_b$ \cite{Varro1,Glauber}. Acting with both of the operators on the vacuum of the original Fock space will define a coherent squeezed state.\nn
In the following, the stationary Schr\"odinger equation is considered because the spectrum is the focus of the present study. It reads
\bea\label{shift}
\mathcal{H} \Psi_{{\bol{p}},n}=E_{{\bf{p}},n}\Psi_{{\bol{p}},n},
\eea
with the energy levels
\bea\label{enlevel}
E_{{\bol{p}},n}=\frac{{\bf p}^2}{2m}+\hbar \Omega\left(n_b+\frac{1}{2}-|\sigma|^2\right),
\eea
and the eigenstates having the form of
\bea
\Psi_{{\bol{p}},n}=\ket{\bol{p}}\otimes D_{\sigma}\ket{n}_b,
\eea 
which is the direct product of the momentum eigenstate of the free charge and the shifted/Bogoliubov-transformed Fock state of the photons \cite{Varro1}. Any state vector of the transformed Fock space can be described by a superposition of state vectors from the original Fock space and thus the two spaces can be considered to be equivalent. The solution of the time dependent Schr\"odinger equation can be found in \cite{Varro1}.

\section{Quasiparticle excitations}\label{dispsec}
Quasiparticles are collective excitations of a given interacting system \cite{quasi}. Inside plasmas or metals, oscillation of the electron density can produce plasma oscillations. The quanta of such oscillations are called plasmons \cite{plasma} and it was shown that for a system of free electron gas, interacting with a coherent LP electromagnetic field can produce such quasi mode \cite{Aryeh}. However, in the current case, only a single charged particle is present; therefore, the charge density is set as $n_e=1/V$. On the other hand, \eqref{startHam} can be modified so that it describes an interaction of $N$ point charges with the electromagnetic field:
\bea
H&=&\frac{1}{2 m}\sum\limits_{i=1}^{N}\left(\bol{p}_i-\frac{e}{c} \bol{A}\right)^2+\hbar\omega\left(\frac{1}{2}+a^\dagger a\right).
\eea
Such a Hamiltonian describes a plasma where the electron-electron interactions were neglected due to the Debye screening, and, further assuming low velocities for the electrons the effects of the electron-ion and electron-atom collisions are also absent \cite{Aryeh}. By using the same method as above, the Hamiltonian can be diagonalized with the following modifications in \eqref{hamil}: ${\bf{p}}^{(2)}\to\sum_i {\bf{p}}_i^{(2)}$ and $e^2\to N e^2$. Apart from these differences, the shape of the Hamiltonian remains the same as in the single-electron case and the eigenstates are the direct product of $N$ free-electron momentum eigenstates and the displaced photon number state with the appropriate modifications in the shift parameter, i.e., $\sigma(\bol{p},e^2)\to\sigma\left(\sum_i\bol{p}_i,N e^2\right)$. In this way, $\omega_p$ can be truly considered as the plasma frequency with charge density of $n_e=N/V$ \footnote{The author thanks to S. Varr\'o the discussions about the many electron scenario and plasma frequency at this point.}. 
The remaining part of the Hamiltonian describes a quantum oscillator with the quasi mode $\Omega$.

\subsection{Dispersion relations}
\begin{figure*}[ht!]
	\centering
	\begin{subfigure}[b]{0.46\textwidth}
		\includegraphics[width=\textwidth]{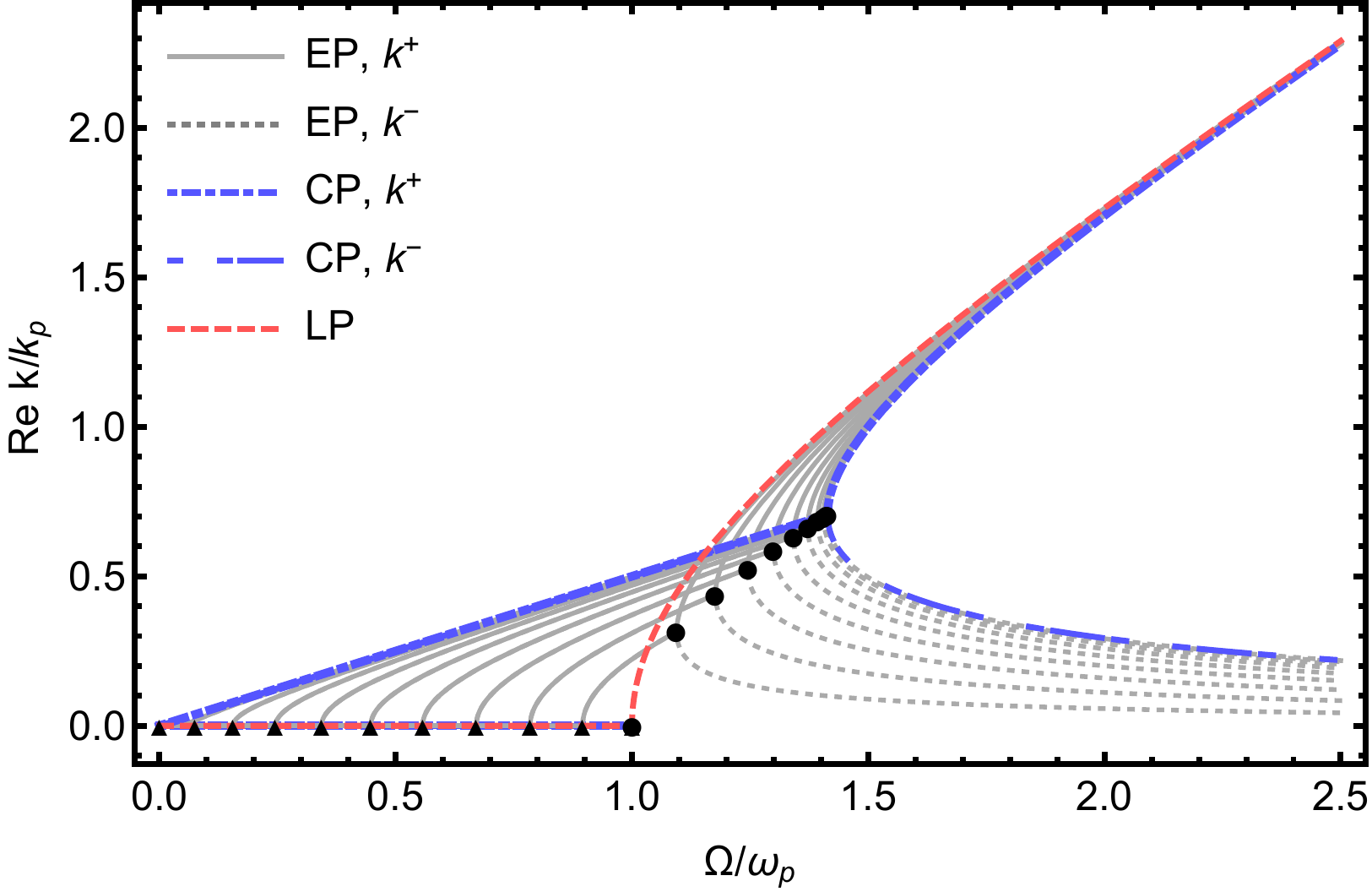}
		\caption{}
		\label{wn:1}
	\end{subfigure}
	\hskip 1cm
	\begin{subfigure}[b]{0.47\textwidth}
		\includegraphics[width=\textwidth]{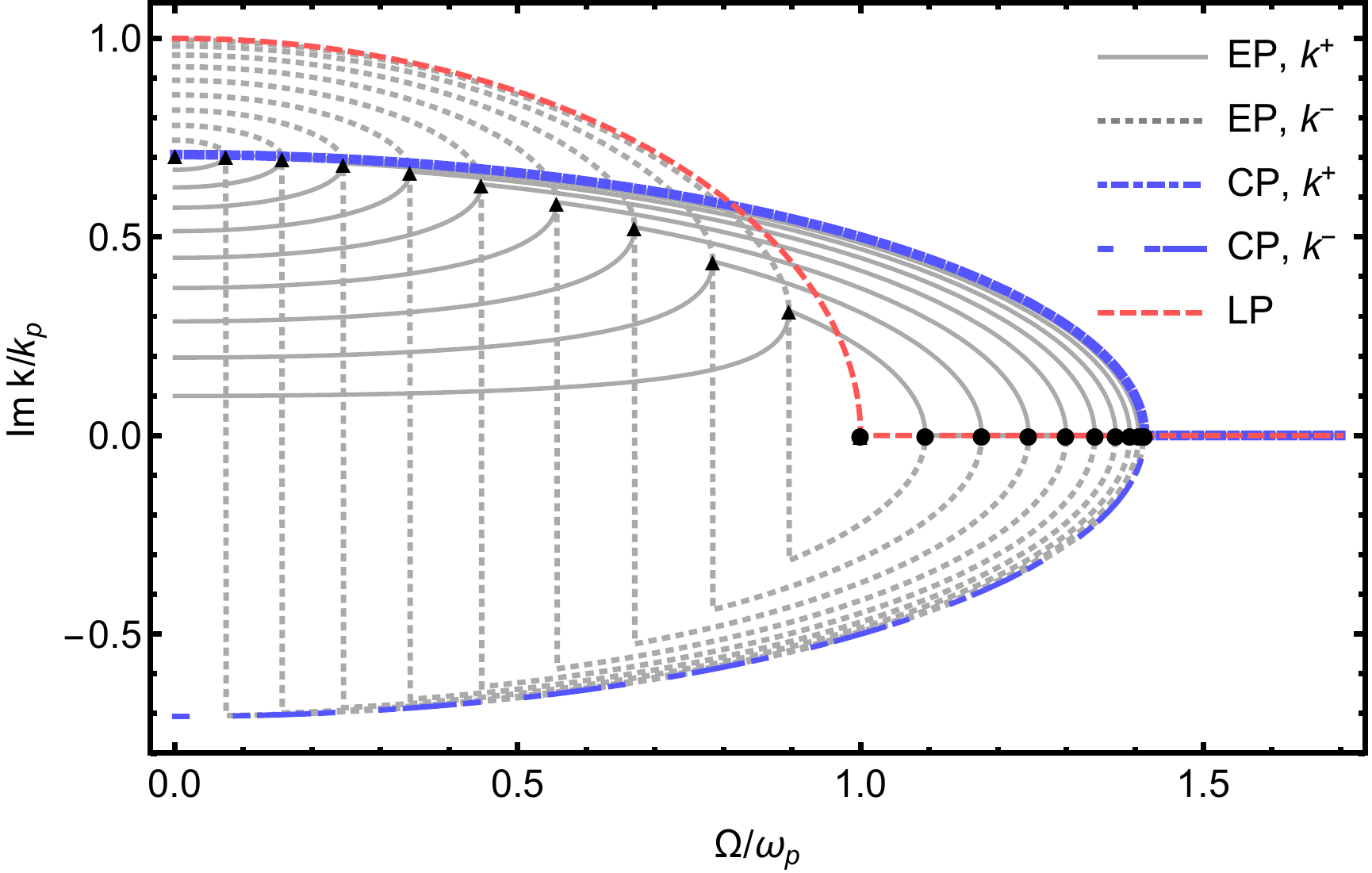}
		\caption{}
		\label{wn:2}
	\end{subfigure}
	\caption{The (a) real and (b) imaginary parts of the wavenumber $k$ for various polarizations. In both panels the $k^+$ and the $k^-$ branches are shown from \eq{wn}. The black dots indicate the value where the wavenumber develops an imaginary part (at $\Omega=\Omega^*$). The black triangles show where the real part of the wavenumber vanishes and hence $k$ becomes purely imaginary (at $\Omega=\tilde{\Omega}$).} 
\end{figure*}

By taking the limit $\bol{p\to 0}$ the system is governed only by the quantum oscillator with frequency $\Omega$. In fact, this limit can be thought of as a uniformly distributed charge in a cube of volume $V$, due to the uncertainty principle in quantum mechanics. In the case of $N$ charges, all $\bol{p}_i$ can be taken to zero, which smears the net charge of $N e$ in the volume $V$. Hence, in this limit the Hamiltonian represents the behavior of EM waves in a plasma where the negative net charge is uniformly distributed. The corresponding dispersion relation can be obtained by using $\omega=c k$, where $k=|\bol{k}|$ is the wavenumber, i.e., the free photon case.
The dispersion relation thus reads
\bea\label{Omega}
\Omega=\sqrt{c^2 k^2+\omega_p^2 \left(1+\frac{\xi^2}{\left(\xi^2+1\right)^2}\frac{\omega_p^2}{c^2 k^2}\right)}.
\eea
\setcounter{equation}{14}
\noindent This expression takes the form for CP and LP, respectively,
\bea\label{omegalinci}\setcounter{equation}{14}
\Omega_{c}= ck\left(1+\frac{1}{2}\frac{\omega_p^2}{c^2 k^2}\right) \text{\, and \,} 
\Omega_{l}=\sqrt{c^2 k^2+\omega_p^2}.
\eea
It is clear that $\Omega_l$ describes a bulk plasmon (or bulk plasmon polariton) \cite{plasma,Aryeh}, with the plasma frequency $\lim_{k\to0}\Omega_l=\omega_{p}$. More precisely, $\Omega_l$ is the dispersion of a transverse electromagnetic wave in a plasma, in contrast to the longitudinal wave which has a constant dispersion $\omega_p$. However, for $\Omega$ in general, i.e., for $0<\xi\leq 1$, the dispersion relation becomes more structured, since besides the constant $\omega_p$, it contains a singular term $\propto 1/(ck)$, too. It is not hard to see that $\Omega$ restores the free photon dispersion at the $V\to\infty$ limit, as $\omega_p$ vanishes. All the functions defined in $\eqref{Omega}$ exhibit a global minimum at  $k^*=k_p\sqrt{\xi /(1+\xi^2)}$, where $k_p=\omega_p/c$.
\begin{figure}[t!]
	\includegraphics[width=0.46\textwidth]{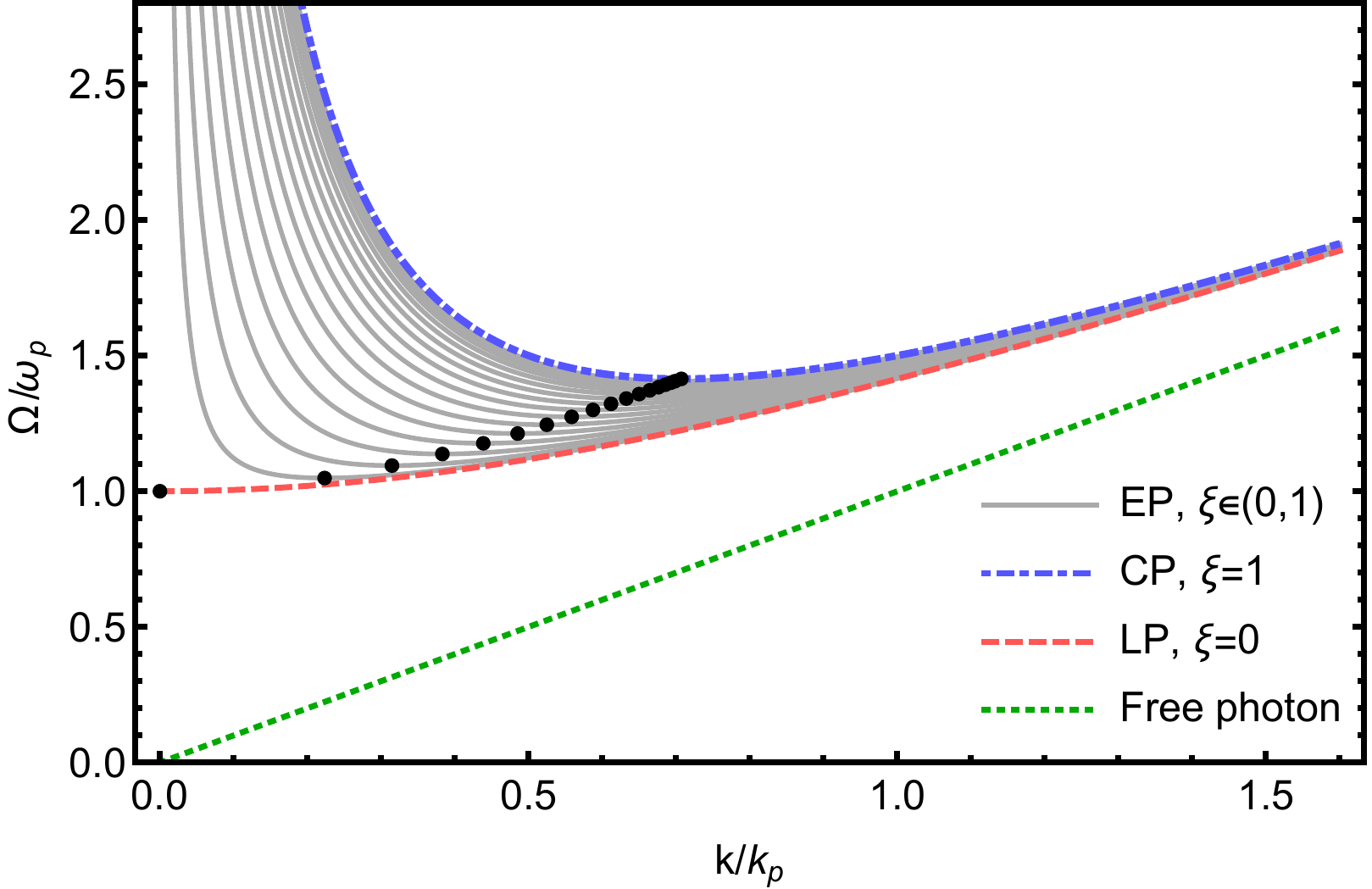}
	\caption{Dispersion relations of the quasi mode $\Omega$ for various polarizations \eqref{Omega}. The frequency and the wavenumber are rescaled by $\omega_p$ and $k_p$, respectively. The black dots represent the minimum of the functions. The dispersion of the free photon is shown for reference.}\label{disper}
\end{figure}
The dispersion relations are shown in \fig{disper}.
The only function from the dispersion relations that has the minimum at $k^*=0$ is for LP and $\Omega_l(0)=\omega_p$. In general, $\Omega(k^*)=\Omega^*= \omega_p\,(1+\xi)/\sqrt{1+\xi^2}$ at the minimum. $\omega_{p}$ is the quantity that determines the frequency below which light waves are fully reflected in the case of plasmas or metals \cite{plasma}. Indeed, considering $\Omega_l$ and expressing $k$, the condition of the solutions for $k\geq 0$ is $\Omega_l\geq\omega_p$. Otherwise the wavenumber takes imaginary values which cannot be associated to any traveling wave. A similar analysis can also be performed for the other polarizations. In general, expressing the wavenumber from the dispersion relation reads
\bea\label{wn}
\! \! \! \! \! \! \! \! \!  k^{\pm}=\frac{1}{\sqrt{2} c}\sqrt{\Omega ^2-\omega_{p} ^2\pm\sqrt{\left(\Omega ^2-\omega_{p} ^2\right)^2-\frac{4  \xi ^2 \omega_{p}^4}{\left(\xi ^2+1\right)^2}}}.
\eea
Except for the LP, where this relation reduces to $k=1/c \sqrt{\Omega^2-\omega_p^2}$, two distinct branches define the wavenumber, denoted by $k^+$ and $k^-$. Strictly speaking, \eqref{wn} with an overall negative sign also gives the right dispersion in \eqref{Omega}, however, it would define negative wavenumbers, which is physically meaningless. In order to discuss properties that are related to wave propagation the dipole approximation of the vector potential might not be satisfactory. However, as it was mentioned earlier, the same dispersion relation can be derived for plane wave vector potential in the $\bol{p\to 0}$ limit (see \app{pwgen}), in a similar manner as it was done in \cite{Varro2}. Thus, by considering a plane wave, the positivity of the real part of the wavenumber indicates a propagating wave in the medium. However, as soon as the imaginary part develops a nonzero value, a damping of the oscillatory wave occurs, corresponding to a finite penetration depth. The real and the imaginary parts of the wavenumber for various polarizations are shown in \fig{wn:1} and \fig{wn:2}, respectively. It is clear from the figures that the above statement about the total reflection is only true for the LP case (dashed red line): For $\Omega<\Omega^*=\omega_p$ the real part of the wavenumber vanishes and the imaginary part becomes finite. For all the other cases the real part remains finite even for values $\Omega<\Omega^*$, and it only disappears at a distinguished value $\tilde{\Omega}=\omega_p\, (1-\xi) /\sqrt{\xi ^2+1}$, where from \eqref{wn}, $k^\pm(\tilde\Omega)=k_p\,\sqrt{-\xi/(\xi^2+1)}$. Thus, for frequencies $\Omega<\tilde{\Omega}$ the wavenumber becomes imaginary just like for the LP, and hence a complete reflection of EM waves is present. Therefore, $\tilde{\Omega}$ can be considered as a modified plasma frequency for polarizations different from LP. In other words, the propagation of EM waves in a plasma highly depends on its polarization. The following statement can be formulated: An EM wave in a plasma is
\begin{itemize}
\item a traveling wave for frequencies $\Omega\geq\Omega^*$, 
\item a decaying traveling wave for $\tilde{\Omega}<\Omega<\Omega^*$,
\item and a decaying standing wave for $\Omega\leq\tilde{\Omega}$ (evanescent wave).
\end{itemize}

\begin{figure}[hb!]
	\includegraphics[width=0.47\textwidth]{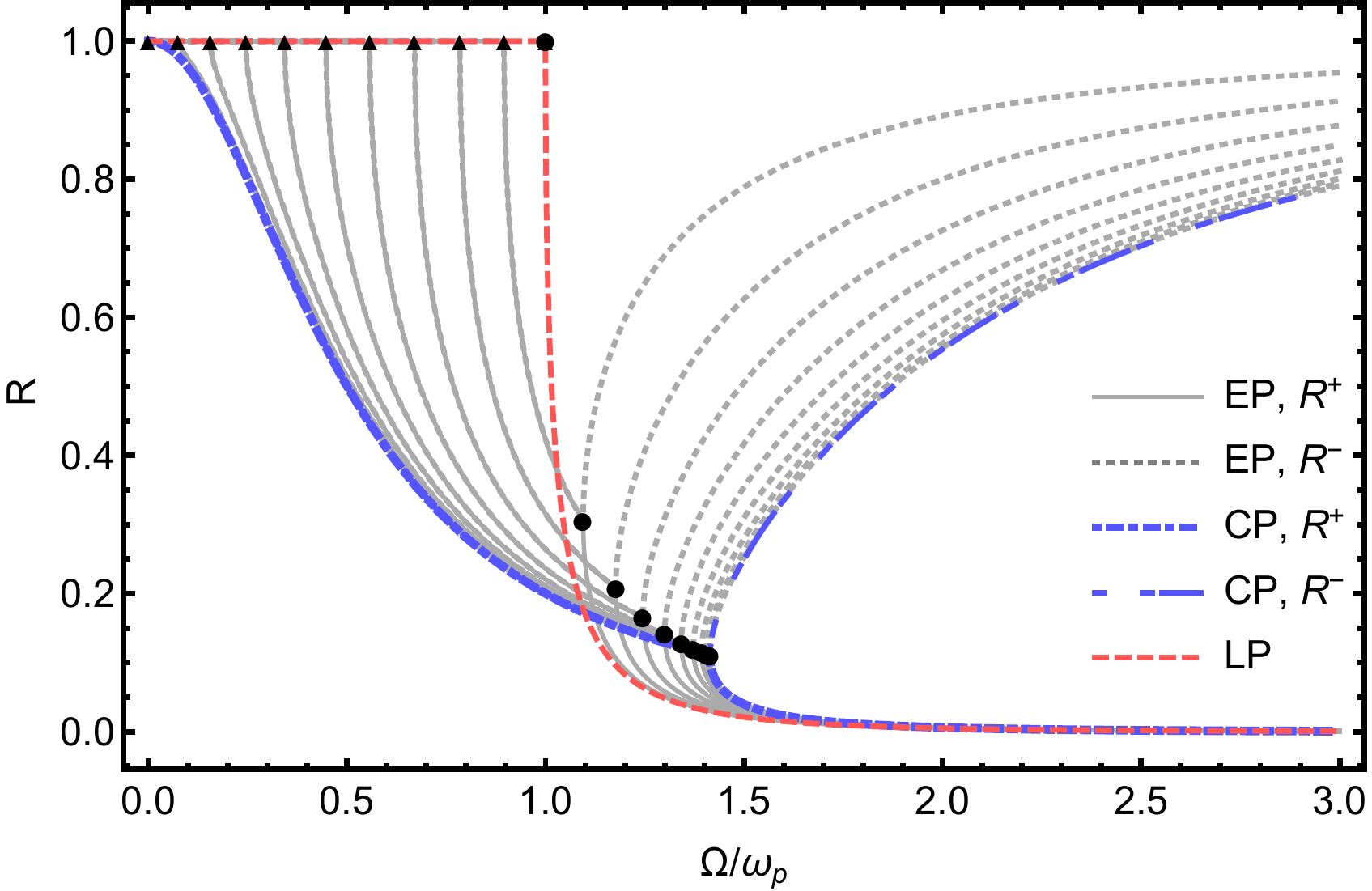}
	\caption{Reflectivity of EM waves in plasma. The well-known curve is obtained for the LP case. For all the other polarizations the complete reflection is at $\tilde{\Omega}$ and the curves bifurcate at $\Omega^*$, likewise in \fig{wn:1}.}\label{refl}
\end{figure}

\noindent \fig{wn:2} shows that for $\Im k$ besides the two branches that develop for $\Omega<\Omega^*$, the $k^-$ branch has a jump discontinuity at $\Omega=\tilde{\Omega}$ from $\Im{k^-}=-k_p \sqrt{\xi/(1+\xi^2)}$ to $\Im{k^-}=k_p \sqrt{\xi/(1+\xi^2)}$, whereas the $k^+$ branch has a cusp. The two branches then bifurcate again at this point and they terminate at $\Omega=0$, where $k^+(0)=k_p\xi/\sqrt{1+\xi^2}$ and $k^-(0)=k_p/\sqrt{1+\xi^2}$.
The CP case behaves somewhat differently. The frequency where the imaginary part develops a finite value is at $\Omega_c^*=\sqrt{2}\omega_p$, and that where the real part vanishes is at $\tilde{\Omega}_c=0$. This means that for CP there is never a total reflection of the EM waves; however, a damping still occurs as the imaginary part of the wavenumber is nonzero for frequencies below $\sqrt{2}\omega_p$. For the CP case there is only one bifurcation of the imaginary part of the wavenumber: The two branches depart from $\Omega=\sqrt{2}\omega_p$ and continue all the way to $\Omega=0$, where $\lim_{\Omega\to0}\Im k^{\pm}=\pm k_p/\sqrt{2}$.\\
In order to give a more detailed insight in the reflectivity property of such a system, the dielectric function (or relative permittivity) must be given
\bea\!\!\!\!\!\!\!\!\!
\zeta^\pm=\frac{1}{2} \left[1\pm\frac{\omega _p^2}{\Omega ^2} \left(\sqrt{\left(1-\frac{\Omega ^2}{\omega _p^2}\right)^2-\frac{4 \xi ^2}{\left(\xi ^2+1\right)^2}}\mp1\right)\!\!\right]\!\!\!.
\eea
The "$\pm$" sign corresponds to the two branches in \eqref{wn}. By taking $\xi\to0$ (LP) the well-known result for the plasma dielectric function is obtained: $\zeta^+=1-\omega_p^2/\Omega^2$ for $\Omega\geq\omega_p$ and $\zeta^-=1-\omega_p^2/\Omega^2$ for $\Omega<\omega_p$, hence $\zeta_l=\zeta^+ \cup \zeta^-= 1-\omega_p^2/\Omega^2$ \cite{plasma}. The refractive index is $\eta=\sqrt{\zeta}$, which can be used to compute the normal incidence reflectivity:
\bea
R=\left|\frac{\eta-1}{\eta+1}\right|^2.
\eea
This function is shown in \fig{refl} for various polarizations. The LP case (dashed red line) shows the reflectivity associated to the well-known plasma relative permittivity: For frequencies $\Omega\leq\omega_p$ the EM waves are reflected completely, i.e., $R=1$. This is in accordance with what \eqref{wn} predicts for the LP case. On the other hand, for all the other polarizations ($0<\xi\leq1$) the reflectivity function behaves differently: As it could be extracted from \eqref{wn}, transmission of the EM wave still occurs below $\omega_p$ and only at $\tilde{\Omega}$ does it reflect completely. Another interesting behavior can be identified at $\Omega^*$: For frequencies $\Omega>\Omega^*$ the reflectivity function bifurcates similarly to \fig{wn:1}. In fact, this clearly occurs as a consequence of the degeneracy of the frequency $\Omega$ in the wavenumber. It is apparent that the waves corresponding to the $k^-$ branch have a higher reflectivity, and as $\Omega\to\infty$ they reflect completely, whereas the reflectivity of the $k^+$ waves falls off rapidly and tends to zero as $\Omega\to \infty$.

\subsection{Phase and group velocities}
It is instructive to study the phase and group velocities of the corresponding quasi mode. The former is obtained by dividing the dispersion by the wavenumber:
\bea
v_{ph}=\frac{\Omega}{k}=c \sqrt{1+\frac{k_p^2}{k^2}+\frac{\xi ^2}{\left(\xi ^2+1\right)^2}\frac{k_p^4}{k^4}}.
\eea
It is apparent that the phase velocity is always greater than the speed of light since $c$ has a factor which is greater than one independently of $k$ and $\xi$. In particular, at the minimum, $v_{ph} ^c(k^*)= 2 c$ (for CP) and $v_{ph}^l(k^*)=\infty$ (for LP).
The latter is characteristic of the longitudinal plasma oscillation. The group velocity is defined as the wavenumber derivative of the dispersion relation:
\bea\label{gv}
v_g=\frac{\partial \Omega}{\partial k}=c\frac{1-\frac{\xi ^2}{ \left(\xi ^2+1\right)^2}\frac{k_p^4}{k^4}}{\sqrt{1+\frac{k_p^2}{k^2}+\frac{\xi ^2}{\left(\xi ^2+1\right)^2}\frac{k_p^4}{k^4}}}.
\eea 
\begin{figure}[ht!]
	\includegraphics[width=0.45\textwidth]{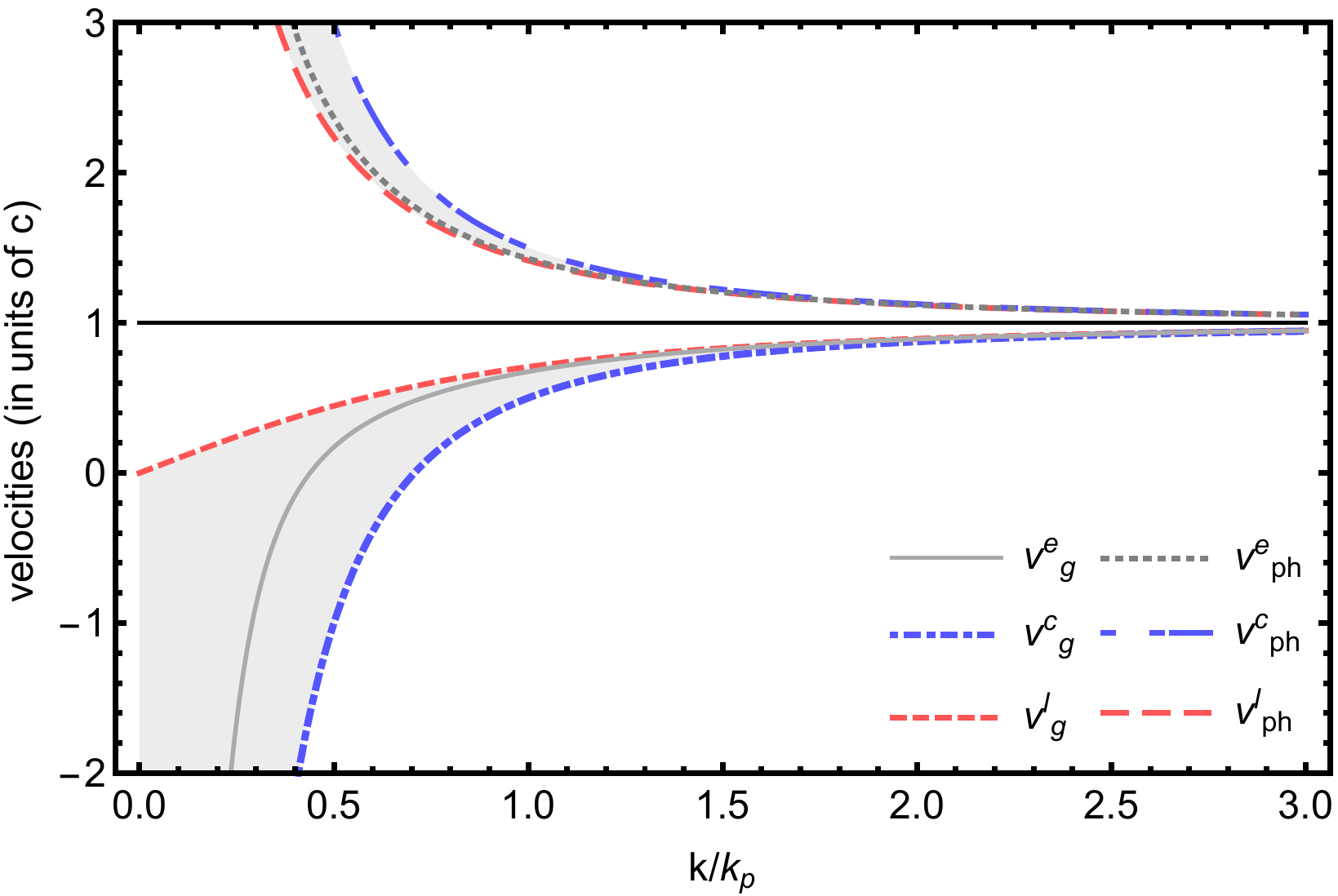}
	\caption{The phase and group velocities above and below light speed (solid black line), respectively. The EP cases are in the shaded area, and likewise for the gray line for which $\xi=0.2$, bounded by the CP and LP velocities.}\label{velo}
\end{figure}

\noindent At the minimum of the dispersion relation the group velocity clearly vanishes, i.e. $v_g(k^*)=0$, describing a localized oscillation for all polarizations. A more interesting observation can be made for non-LP cases below their $k^*=k_p\sqrt{\xi/(1+\xi^2)}$ minimum: As the dispersion relation is a decreasing function of the wavenumber in this region, the group velocity exhibits negative values. Moreover, its value even can exceed the speed of light in absolute value. Note that, in this situation both the group and the phase velocities are larger than the light speed, but their orientation is opposite. In particular for the CP, this threshold is $k=k_p/2$, and for EP it is $k=k_p\,\xi ^{2/3} \sqrt{-\xi ^{2/3}+\xi ^{4/3}+1}/\left(\xi ^2+1\right)$. Even though this seems to violate causality, in fact, the group velocity cannot be identified with the propagation speed of the information when the dispersion is anomalous \cite{Brilluin,Boyd1}. Experimental observations of "fast" and backward-propagating light pulses are reported in \cite{FastLight}. For the LP no negative group velocity is observed, since the numerator in \eqref{gv} cannot be negative for $\xi=0$. The phase and group velocities as a function of the wavenumber for various polarizations are shown in \fig{velo}.

\section{Energy spectra and repulsive force}

\begin{figure*}[ht!]
	\centering
	\begin{subfigure}[b]{0.4\textwidth}
		\includegraphics[width=\textwidth]{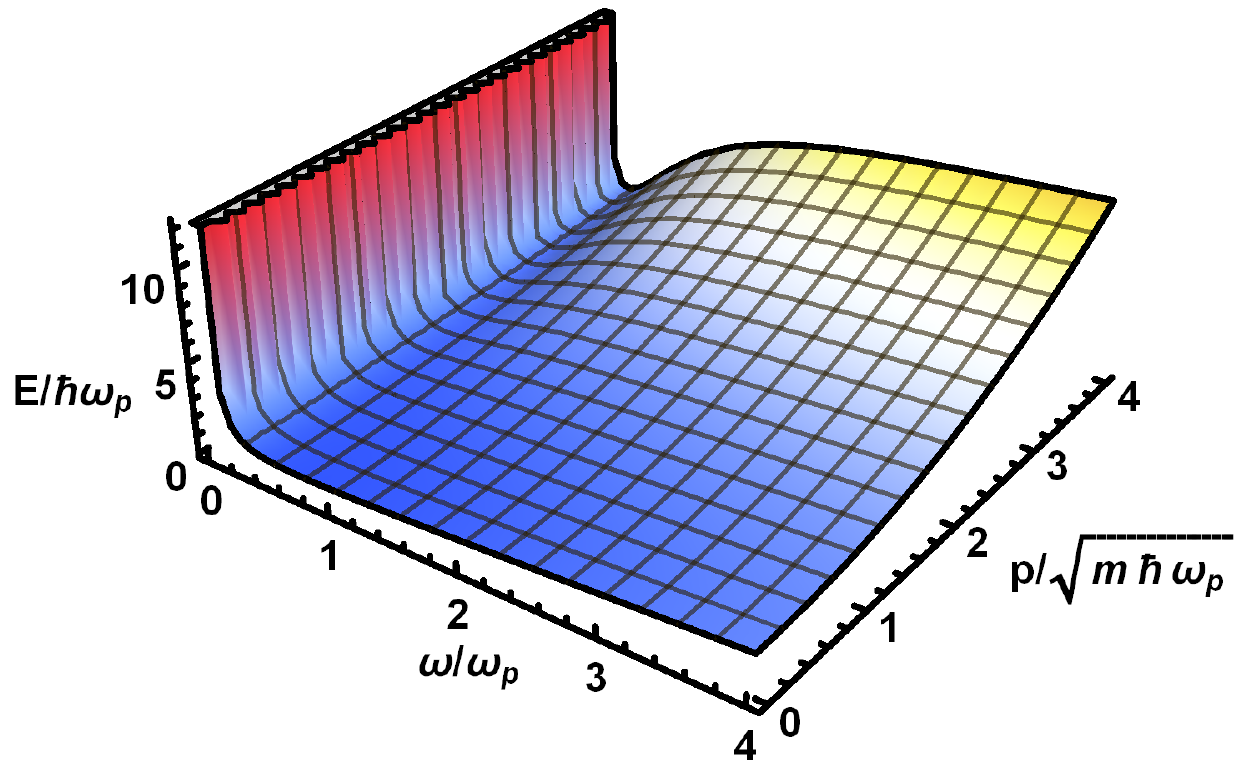}
		\caption{}
		\label{fig:1}
	\end{subfigure}
	\hskip 1cm
	\begin{subfigure}[b]{0.4\textwidth}
		\includegraphics[width=\textwidth]{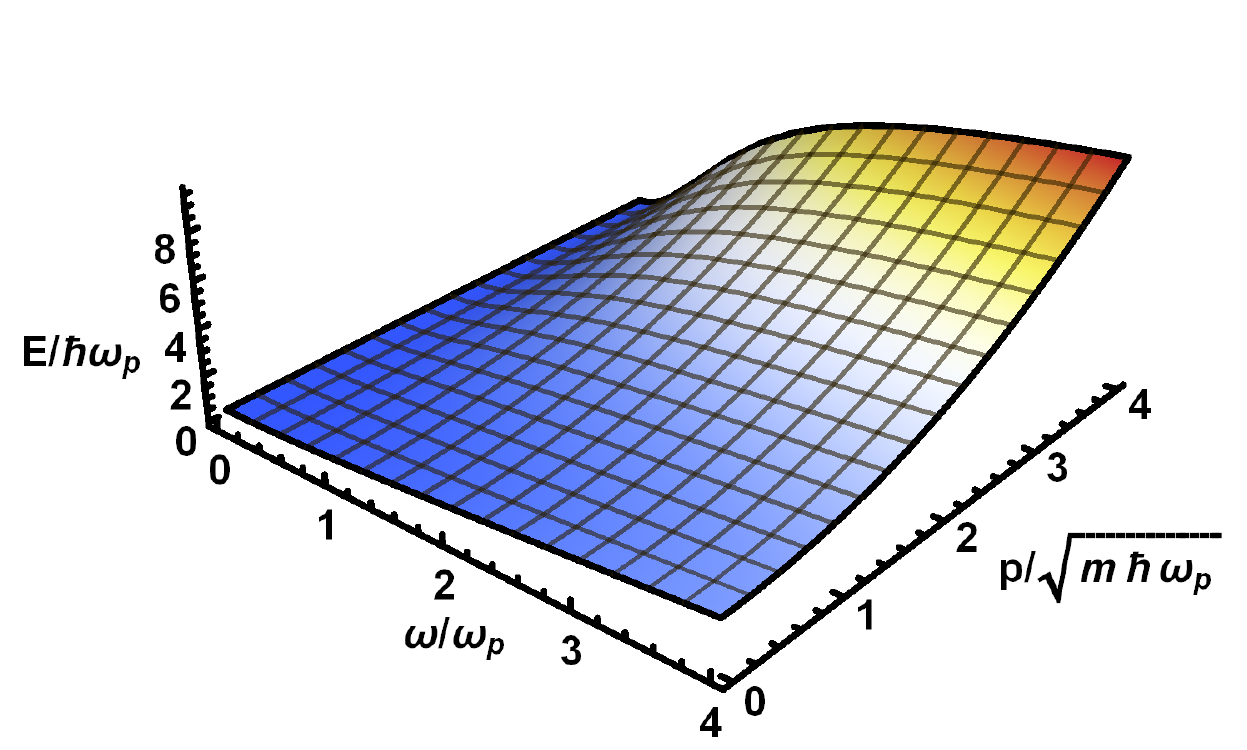}
		\caption{}
		\label{fig:2}
	\end{subfigure}
	\caption{The energy spectrum of (a) the CP and (b) the LP case in a $(p,\omega)$ plot, with the oscillator excitation number set to $n_{a(b)}=0$. The momentum is chosen so that in \eqref{energies} (a) $p_z=0$ in $E_c$ and (b) $\bf{p}$ is parallel to $\bsy\epsilon$ and hence $\phi=0$ in $E_l$. The most apparent difference between the two plot is in the $\omega\to0$ limit: (a) exhibiting a singular behavior and (b) having a finite value $E_l=\hbar \omega_p/2$. The axes have been rescaled appropriately for clarity.}\label{EnCL}
\end{figure*}

In the following, only the energy spectrum of the CP and LP cases are analyzed in full details. Evaluating \eqref{enlevel} for these two separate cases
\bea\label{energies}
E_c&=& \frac{2 p^2 \omega^2+p_z^2 \omega_{p}^2}{2 m \left(2 \omega ^2+\omega_{p}^2\right)}+\hbar \omega \left(1+\frac{\omega _{p}^2}{2 \omega^2 } \right)\left(\frac{1}{2}+n_a\right),\\
E_l&=&\frac{p^2}{2 m} \left(1-\frac{\omega _{p}^2 \cos ^2(\phi )}{\omega ^2+\omega _{p}^2}\right)+ \hbar  \sqrt{\omega ^2+\omega _{p}^2}\left(\frac{1}{2}+n_b\right),\nonumber
\eea 
where $\phi$ in $E_l$ is the angle between the momentum ${\bf{p}}$ and the polarization vector $\bsy\epsilon$. Like the dispersion relation $\Omega_c$, the energy spectrum $E_c$ also diverges as $\omega\to0$ -- a characteristic for all non-LP cases. For every fixed $\bf p$ and $n_a$ value a minimum of the energy can be found with respect to $\omega$ which for $\bf{p}\to0$ with $n_a=0$ gives $E_c^*=\hbar\Omega_c^*/2=\hbar \omega_p/\sqrt{2}$. This can be considered as the lowest value of the zero-point energy of the system. In general $E^*=\hbar\omega_p(1+\xi)/2\sqrt{1+\xi^2}$, which of course reduces to $E_l^*=\hbar\omega_p/2$ in the case of the LP light. Looking at $E_l$ in \eqref{energies}, an interesting case can be observed, when ${\bf p}$ is parallel with the polarization vector $\bsy\epsilon$. Considering this situation with $\omega\to0$ the first term of $E_l$ yields zero. This means that independently of the momentum $\bf p$, the energy of the system is $E_l^*=\hbar\omega_p/2$. The energy spectrum as a function of the momentum $p=|{\bf{p}}|$ and $\omega$ for the CP and LP cases are shown in \fig{EnCL}. In both panels the excitation number $n_{a(b)}=0$; however, the character of the plot would have the same features for finite $n_{a(b)}$ values, too. For the non-LP cases, besides the singular nature, the degeneracy is also transferred to the energy spectrum from the dispersion relation, meaning that two distinct frequencies $\omega^{\pm}(=c k^{\pm})$ are associated to a particular energy level above the corresponding minimum.\par
The presence of the zero-point energy is the most significant difference between the quantized system presented above and its classical equivalent -- where the operators become $c$-numbers and hence the Hamiltonian becomes classical. The zero-point energy minimum can be rewritten as $E^*=\kappa \hbar e \sqrt{\pi/m V}$, with $\kappa\equiv(1+\xi)/\sqrt{1+\xi^2}$. From this form it is clear that the finite $E^*$ is a consequence of the finite quantization volume, i.e., it vanishes in the $V\to\infty$ limit, and the zero-point energy becomes the usual expression of the free quantum harmonic oscillator, i.e., $E=\hbar \omega/2$. There is another well-known finite-volume phenomenon, the Casimir effect, that deals with the zero-point energy \cite{Casimir,Milton,Schwartz}. In contradistinction to the current monochromatic case, however, in order to obtain the Casimir effect, all the modes of the EM field must be summed, resulting in a divergent series $E_{\text{Cas}}=\sum_k \hbar \omega_k/2$, which can be evaluated by using appropriate regularization schemes (see, e.g., \cite{Schwartz}). In this case the zero-point energy implies an attractive force between two parallel perfectly conductive plates with a surface area $\mathcal{A}$. The result was derived first by Casimir \cite{Casimir}; he obtained the force per unit area $\mathcal{F}_{\text{Cas}}\equiv F_{\text{Cas}}/\mathcal{A}=-\hbar c \pi^2 /240\,d^4$, where $d$ is the distance between the plates. The same lines of thought can be applied to the present model, where the finite-volume dependence is carried by the plasma frequency $\omega_p(V)$, and the volume of the box (in which a total number of $e$ charge is smeared uniformly) is defined with the parameters of the two-plate layout ($V=\mathcal{A}\,d$), likewise in the Casimir effect experiment. The force between the two plates is defined as the derivative of the negative energy with respect to their distance, i.e., $F=-\partial E/\partial d$. Unlike in the case of the Casimir effect, the current force is frequency dependent due to the lack of mode summation. However, the minimum of the zero-point energy can be considered as a distinguished point and hence the force for this special case is presented here. The general case with frequency dependence, as well its derivation, can be found in \app{forceapp}. The force implied by the minimum of the zero-point energy reads
\setcounter{equation}{20}
\bea\label{force}
F^*=\frac{\kappa}{2}\sqrt{\frac{\pi R_B}{\mathcal{A}}} \frac{e^2}{d^{3/2}},\qquad 
\mathcal{F^*}=\frac{\kappa}{2} \frac{\sqrt{\pi R_B}}{{\mathcal{A}}^{3/2}} \frac{e^2}{d^{3/2}},
\eea 
where $R_B=\hbar^2/m e^2$ was introduced as the Bohr radius and $\kappa\in[1,\sqrt{2}]$. $F^*$ represents a repulsive force, scaling with the distance as $\propto d^{-3/2}$. Moreover, the force per unit area $\mathcal{F^*}$, unlike $\mathcal{F}_{\text{Cas}}$, is not independent of the surface area: The repulsion of the plates increases as the surfaces decrease. Tuning the characteristic length scale of the system to the order of the Bohr radius, i.e., $d\sim\sqrt{\mathcal{A}}\sim R_B$, the force behaves as $F^*\propto e^2/R_B^2$.\\
Alternatively, it is possible to express the force by using the plasma frequency
\bea
F^*= {\mathcal{F}}^* {\mathcal{A}}=\frac{\kappa}{4} \frac{\hbar\,\omega_p(d,{\mathcal{A}})}{d}\approx\frac{\kappa}{4} \frac{\hbar\, \omega_p}{d}.
\eea
However, when using this representation, it must be borne in mind that the plasma frequency itself depends on the distance and the surface area, which is indicated in the argument $\omega_p(d,{\mathcal{A}})$. The simplified scaling of $F^*\propto d^{-1}$ is valid only when the $\Delta d$ change in the distance makes a negligible correction to the plasma frequency, i.e., $\omega_p(d+\Delta d)\approx \omega_p$. It is apparent that the implied force depends on the charge, too. In fact, no such repulsive force would be present if the charge vanished, since $E^*\to0$ as $e\to0$. On the other hand, the Casimir force seemingly does not depend on the charge at all. However, Jaffe showed the contrary in \cite{Jaffe}: In his argument the Casimir-force originates from the interaction between the EM modes and the conducting plates; thus, it can be shown that it also vanishes as $e\to0$. In that sense, this property agrees with the findings about the present repulsive force. By replacing the vacuum with appropriate dielectric materials, Lifshitz \textit{et al.} in \cite{Lifshitz} showed that a repulsive type of Casimir-force can be achieved. This effect was experimentally verified by Munday \textit{et al.} \cite{Munday}.\\
The above results equally hold the for $N e$ charges and $n$ (quasi)photons (i.e., non-vacuum states) that could enhance the magnitude of the force:
\bea
F^*=\kappa \sqrt{\frac{\pi R_B}{\mathcal{A}}} \frac{N e^2}{d^{3/2}}\left(\frac{1}{2} + n \right).
\eea
It could be of interest to measure such a force in an appropriate experimental setup.

\section{Conclusion}
In summary, a non-relativistic quantum mechanical model has been considered consisting of a charged particle interacting with one electromagnetic radiation mode. The diagonalization of the Hamiltonian leads to plasmon-like quasiparticle excitations from which modified plasma frequencies ($\tilde{\Omega}\leq\omega_p$) can be found for the non-LP cases. The reflectivity function also shows a strong dependence on the polarization of the EM wave: total reflection can only be found for frequencies $\Omega<\tilde{\Omega}$. For instance, in the case of CP, there is no complete reflection of the wave at any finite frequency since $\tilde{\Omega_c}=0$.
The phase and group velocities are also determined for different polarizations. It is found that the phase velocity of the quasi modes always exceeds the light speed, independently of the polarization. The group velocities vanish at the position of the minimum of the dispersions which for the LP coincides with zero wavenumber. For non-LP cases, the group velocities can take negative values and even exceed the light speed in absolute value. The zero-point energy of the system is derived which implies a repulsive force between two parallel plates. The force scales as $\propto 1/d$ with the distance between the plates when the change in the plasma frequency, due to the volume increasement, is negligible. If the change in the plasma frequency is considerable with respect to the volume growth, the force, at the minimum of zero-point energy, scales as $\propto d^{-3/2}$.

\section*{ACKNOWLEDGMENTS}
The author thanks S. Varr\'o, A. Jakov\'ac, H. Gies, M. Horv\'ath, and G. S\'arosi for useful discussions and comments. The ELI-ALPS project (GINOP-2.3.6-15-2015-00001) is supported by the European Union and co-financed by the European Regional Development Fund.

\appendix
\section{Diagonalization of the Hamiltonian for various polarizations}\label{app1}

In the following the details of the diagonalization of the Hamiltonian 
\bea\label{startHam2}
H&=&\frac{1}{2 m}\left(\bol{p}-\frac{e}{c} \bol{A}\right)^2+\hbar\omega\left(\frac{1}{2}+a^\dagger a\right)
\eea 
are presented for an elliptically polarized field along the lines applied in \cite{Varro1} for the circularly polarized and linearly polarized cases. Here 
\bea\label{elli2}
\bol{A}_e=\alpha(\mathbfcal{E} \,a+\mathbfcal{E}^{*}\,a^\dagger),
\eea
where $\mathbfcal{E}$ must be parametrized so that in certain limits it gives the polarization vectors of the CP and LP cases, i.e., $\mathbfcal{E}(\xi\to1)=\bsy\varepsilon$ and $\mathbfcal{E}(\xi\to0)=\bsy\epsilon$ with some parameter $\xi\in[0,1]$. In particular, the parametrization $\mathbfcal{E}=\sqrt{1/(1+\xi^2)}(\Re\bol{u}+i\xi\Im\bol{u})$ with $\bol{u}=(1,i)$  gives the right limits and its norm $\mathbfcal{E}\mathbfcal{E}^*=1$, independently of $\xi$.
The Hamiltonian will have the following form after substituting \eqref{elli2}:
\begin{eqnarray}\label{dahamil}
\begin{aligned}
H=&\frac{\bol{p}^2}{2 m}+\hbar\omega\left(\frac{1}{2}+a^\dagger a\right)-\frac{e\alpha}{m c}\bol{p}(\mathbfcal{E} a+ \mathbfcal{E}^*a^\dagger)\\
&+\frac{e^2 \alpha^2}{2 m c^2}(\mathbfcal{E} a+ \mathbfcal{E}^*a^\dagger)^2\\
=&
\frac{\bol{p}^2}{2 m}-
\frac{e\alpha}{m c}\bol{p}(\mathbfcal{E} a+ \mathbfcal{E}^*a^\dagger)\\
&+\left[\hbar\omega + \frac{e^2 \alpha^2}{mc^2}\right]
\left(\frac{1}{2}+a^\dagger a\right)\\
&+\frac{e^2 \alpha^2}{2 m c^2}\frac{1-\xi^2}{1+\xi^2}\left[a^2 + (a^\dagger)^2\right].
\end{aligned}
\end{eqnarray}
The $\bol{p}$-independent part of the Hamiltonian can be rewritten in terms of the operators $b$ and $b^\dagger$ obtained by Bogoliubov transformation:
\bea\label{railo}
\!\!\!\!\!\! b=\cosh {\theta} a + \sinh{\theta} a^\dagger, \quad
b^\dagger=  \cosh{\theta} a^\dagger + \sinh {\theta} a.
\eea
The Hamiltonian reads
\setcounter{equation}{8}
\begin{eqnarray}
H&=&\frac{\bol{p}^2}{2 m}-\frac{e\alpha}{m c}\bol{p}(\mathbfcal{E} a+ \mathbfcal{E}^*a^\dagger)\nn
&&+\left[\hbar\omega + \frac{e^2 \alpha^2}{mc^2}\right]\nn
&&\times\left[-\frac{\sinh 2 \theta}{2}\left(b^2+(b^\dagger)^2\right)+ \cosh 2 \theta b^\dagger b + \sinh^2 \theta +\frac{1}{2} \right]\nn
&&+\frac{e^2 \alpha^2}{2 m c^2}\frac{1-\xi^2}{1+\xi^2}\nn
&&\times\left[\cosh2\theta \left(b^2+(b^\dagger)^2\right)-\sinh 2 \theta (2 b^\dagger b +1) \right].
\end{eqnarray}
In order to eliminate the quadratic terms in $b^{(\dagger)}$ the following definition for $\theta$ is required:
\setcounter{equation}{5}
\bea
\label{thet}
\tanh 2 \theta = \frac{e^2 \alpha^2}{ m c^2}\frac{1-\xi^2}{1+\xi^2}\frac{1}{\hbar\omega + \frac{e^2 \alpha^2}{mc^2}}.
\eea
Concerning the CP and LP the following expressions for $\theta$ can be found:
\bea\label{thetaxi}
\begin{aligned}
	\left. \tanh 2 \theta \right|_{\xi=1}& = 0,\\
	\left. \tanh 2 \theta \right|_{\xi=0}&=\frac{e^2 \alpha^2}{m c^2} \frac{1}{\hbar\omega + \frac{e^2 \alpha^2}{mc^2}}.
\end{aligned}
\eea
For the CP case [the first equation in \eqref{thetaxi}], it also means that the Bogoliubov transformation is trivial, i.e., $b^{(\dagger)}=a^{(\dagger)}$. Thus, what remained from the Hamiltonian is the following expression:
\bea
\begin{aligned}
	H=&\frac{\bol{p}^2}{2 m}-\frac{e\alpha}{m c}\bol{p}(\mathbfcal{E} a+ \mathbfcal{E}^*a^\dagger)\\
	&+\left[\hbar\omega + \frac{e^2 \alpha^2}{mc^2}\right]\\
	&\times\left[\cosh 2 \theta b^\dagger b + \sinh^2 \theta +\frac{1}{2} \right]\\
	&+\frac{e^2 \alpha^2}{2 m c^2}\frac{1-\xi^2}{1+\xi^2}\\
	&\times\left[-\sinh 2 \theta (2 b^\dagger b +1) \right].
\end{aligned}
\eea
After some algebra and using the hyperbolic function identities,
\bea
\begin{aligned}
	H&=\frac{\bol{p}^2}{2 m}-\frac{e\alpha}{m c}\bol{p}(\mathbfcal{E} a+ \mathbfcal{E}^*a^\dagger)\\
	&+\frac{e^2 \alpha^2}{ m c^2}\frac{1-\xi^2}{1+\xi^2}\frac{1}{\sinh 2 \theta}\left[ b^\dagger b + \frac{1}{2}\right].
\end{aligned}
\eea
As the quadratic terms have been eliminated, the focus can be shifted to the linear terms which also contain the momentum dependence.\\
The second term in the Hamiltonian that consists of linear terms in the creation and annihilation operators can be rewritten in terms of $b$ and $b^\dagger$, giving
\setcounter{equation}{11}
\begin{eqnarray}
&&\frac{\alpha e}{m c} \frac{{\bol{p}}}{\sqrt{(1+\xi^2)}}\left[(\Re\bol{u}+i\xi\Im\bol{u})a + (\Re\bol{u}-i\xi\Im\bol{u})a^\dagger \right]\nn
&&=\frac{\alpha e}{m c} \frac{{\bol{p}}}{\sqrt{(1+\xi^2)}}\left[e^{-\theta}(b+b^\dagger)\Re\bol{u}+e^{\theta}(b-b^\dagger)i \xi \Im\bol{u}\right].\nn
\end{eqnarray}
\setcounter{equation}{11}
After this transformation the Hamiltonian reads
\beq
\begin{aligned}
	H=&\frac{\bol{p}^2}{2 m}-\frac{\alpha e}{m c}\frac{{\bol{p}}}{\sqrt{(1+\xi^2)}}\\
	&\times\left[e^{-\theta}(b+b^\dagger)\Re\bol{u}+e^{\theta}(b-b^\dagger)i \xi \Im\bol{u}\right]\\
	&+ \frac{e^2 \alpha^2}{ m c^2}\frac{1-\xi^2}{1+\xi^2}\frac{1}{\sinh 2 \theta}\left[ b^\dagger b + \frac{1}{2}\right].
\end{aligned}
\eeq
For the elimination of the linear terms the displacement operator is used:
\bea
D_\sigma=\exp(\sigma b^\dagger-\sigma^\dagger b),
\eea
with $\sigma$ being arbitrary at this point, but requiring
\bea
[\sigma,\sigma^{(\dagger)}]=[\sigma,b^{(\dagger)}]=[\sigma,\bol{p}]=0.
\eea
Acting on $b$ and $b^\dagger$ adds a shift to the operator,
\bea
D_{\sigma}^{-1} b^{(\dagger)} D_{\sigma} &=& b^{(\dagger)} + \sigma^{(\dagger)}.
\eea
The transformation must also be unitary:
\bea
D_{\sigma}^{-1}=D_{\sigma}^\dagger,
\eea
hence
\bea
D_{\sigma}^{-1}b^\dagger b D_{\sigma}=b^\dagger b+\sigma^\dagger b + b^\dagger \sigma + \sigma^\dagger \sigma.
\eea
Applying the transformation to the Hamiltonian yields
\setcounter{equation}{20}
\bea
H&=&\frac{\bol{p}^2}{2 m} + \frac{e^2 \alpha^2}{ m c^2}\frac{\frac{1-\xi^2}{1+\xi^2}}{\sinh 2 \theta}\left[ b^\dagger b + \frac{1}{2}+\sigma^\dagger \sigma\right]\nn
&&+\frac{e^2 \alpha^2}{ m c^2}\frac{\frac{1-\xi^2}{1+\xi^2}}{\sinh 2 \theta}\sigma^\dagger b
+ \frac{e^2 \alpha^2}{ m c^2}\frac{\frac{1-\xi^2}{1+\xi^2}}{\sinh 2 \theta}b^\dagger \sigma\nn
&& -
\frac{\alpha e}{m c} \frac{{\bol{p}}}{\sqrt{(1+\xi^2)}}\left[e^{-\theta}\Re\bol{u}+e^{\theta}i \xi \Im\bol{u}\right]
(b+\sigma)\nn
&&-
\frac{\alpha e}{m c} \frac{{\bol{p}}}{\sqrt{(1+\xi^2)}}\left[e^{-\theta}\Re\bol{u}-e^{\theta}i \xi \Im\bol{u}\right] (b^\dagger + \sigma^\dagger).\nn
\eea
At this point, the parameter $\sigma$ must be defined so that the linear terms cancel out. Therefore, the following relationship must hold:
\setcounter{equation}{18}
\bea
\frac{e^2 \alpha^2}{ m c^2}\frac{\frac{1-\xi^2}{1+\xi^2}}{\sinh 2 \theta}\sigma=\frac{\alpha e}{m c}\frac{{\bol{p}}}{\sqrt{(1+\xi^2)}}\left[e^{-\theta}\Re\bol{u}-e^{\theta}i \xi \Im\bol{u}\right].\nn
\eea 
This sets $\sigma$ to
\setcounter{equation}{19}
\bea\label{sigma1}
\sigma=\frac{c}{e \alpha}\frac{\sinh 2 \theta}{\frac{1-\xi^2}{1+\xi^2}} \frac{{\bol{p}}}{\sqrt{(1+\xi^2)}}\left[e^{-\theta}\Re\bol{u}-e^{\theta}i \xi \Im\bol{u}\right],\nn
\eea
or equivalently
\setcounter{equation}{20}
\bea
\sigma= \frac{\cosh 2\theta}{\hbar\omega + \frac{e^2 \alpha^2}{mc^2}}\frac{\alpha e}{m c} \frac{{\bol{p}}}{\sqrt{(1+\xi^2)}}\left[e^{-\theta}\Re\bol{u}-e^{\theta}i \xi \Im\bol{u}\right].\nn
\eea
The parameter $\sigma$ for the CP and LP cases can be obtained by taking the limits $\xi\to1$ and $\xi\to0$, respectively:
\setcounter{equation}{21}
\bea
\lim\limits_{\xi\to1} \sigma &=& \frac{\alpha e}{m c} {\bol{p}}{\bsy{\varepsilon}}^*\frac{1}{\hbar\Omega_c}=\sigma_c,\nn
\lim\limits_{\xi\to0} \sigma &=& \frac{\alpha e}{m c} {\bol{p}}{\bol{e}}e^{-\theta}\frac{1}{\hbar\Omega_l}=\sigma_l.
\eea
Here, the effective frequencies are defined for the two separate cases:
\setcounter{equation}{21}
\bea\label{omx}
\Omega_{c}= \omega\left(1+\frac{1}{2}\frac{\omega_p^2}{\omega^2}\right),\quad \Omega_{l}=\sqrt{\omega^2+\omega_p^2},
\eea
and the polarization vectors are
${\bsy\epsilon}=\Re\bol{u}$, ${\bsy{\varepsilon}}=\Re\bol{u}+i \Im\bol{u}$, and $\omega_p^2=4\pi e^2/mV$. The Hamiltonian now approaches its final form:
\bea
H&=&\frac{\bol{p}^2}{2 m} + \frac{e^2 \alpha^2}{ m c^2}\frac{\frac{1-\xi^2}{1+\xi^2}}{\sinh 2 \theta}\left[ b^\dagger b + \frac{1}{2}+\sigma^\dagger \sigma\right]\\
&& -
\frac{\alpha e}{m c} \frac{{\bol{p}}}{\sqrt{(1+\xi^2)}}\left[e^{-\theta}\Re\bol{u}+e^{\theta}i \xi \Im\bol{u}\right]\sigma\nn
&&-
\frac{\alpha e}{m c} \frac{{\bol{p}}}{\sqrt{(1+\xi^2)}}\left[e^{-\theta}\Re\bol{u}-e^{\theta}i \xi \Im\bol{u}\right] \sigma^\dagger.\nonumber
\eea
The third and fourth terms can be rewritten by using the definition of $\sigma$ in \eqref{sigma1},
\setcounter{equation}{25}
\bea
H&=&\frac{\bol{p}^2}{2 m} + \frac{e^2 \alpha^2}{ m c^2}\frac{\frac{1-\xi^2}{1+\xi^2}}{\sinh 2 \theta}\left[ b^\dagger b + \frac{1}{2}+\sigma^\dagger \sigma\right]\nn
&&-2\frac{e^2\alpha^2}{ mc^2}\frac{\frac{1-\xi^2}{1+\xi^2}}{\sinh 2 \theta}\sigma^\dagger \sigma\nn
&=&\frac{\bol{p}^2}{2 m} + \frac{e^2 \alpha^2}{ m c^2}\frac{\frac{1-\xi^2}{1+\xi^2}}{\sinh 2 \theta}\left[ b^\dagger b + \frac{1}{2}-\sigma^\dagger \sigma\right],
\eea
or equivalently, by using the relation in \eqref{thet}
\bea
H&=&\frac{\bol{p}^2}{2 m} + 
\frac{\hbar\omega + \frac{e^2 \alpha^2}{mc^2}}{\cosh 2\theta}\left[ b^\dagger b + \frac{1}{2}-\sigma^\dagger \sigma\right].
\eea
After some algebra the following expression for the effective angular frequency can be found:
\bea\label{omegae}
\Omega(\xi)=\sqrt{\omega ^2+\omega_p^2 \left(1+\frac{\xi^2}{\left(\xi^2+1\right)^2}\frac{\omega_p^2}{\omega^2}\right)},
\eea
where $\omega_p^2=4 \pi e^2/ m V $. Hence the Hamiltonian reads
\bea\label{hamilt}
\mathcal{H}&=&\frac{\bol{p}^2}{2 m} + \hbar \Omega(\xi) \left[ b^\dagger b + \frac{1}{2}-\sigma^\dagger \sigma\right].
\eea
(Here $\mathcal{H}$ is used for the final form of the transformed Hamiltonian in order to distinguish it from the original $H$.)
Taking the limits $\xi\to 1$ and $\xi\to 0$, the Hamiltonians for the CP and LP cases are obtained, with $\Omega_c$ and $\Omega_l$, respectively. These coincide with the results in \cite{Varro1}.

\section{Extension to plane waves}\label{pwgen}
In the following, it is shown that by using an appropriate unitary transformation, as applied in \cite{Varro2}, the Hamiltonian with a plane wave vector potential becomes equivalent to the one with a dipole approximation up to a term of order $\mathcal{O}(\hbar^2)$ when the limit ${\bf{p}}\to{\bf{0}}$ is taken. The vector potential is defined as
\bea
\bol{A}_e=\alpha \left(\mathbfcal{E} \,a e^{i({\bf{kr}}-\omega t)} +\mathbfcal{E}^{*}\,a^\dagger e^{-i({\bf{kr}}-\omega t)} \right).
\eea
In this case the Hamiltonian in \eqref{startHam} becomes 
\begin{eqnarray}\label{pwHamil}
\begin{aligned}
H_{pw}&=
\frac{\bol{p}^2}{2 m}-
\frac{e\alpha}{m c}\bol{p}\left(\mathbfcal{E} a e^{i({\bf{kr}}-\omega t)} + \mathbfcal{E}^*a^\dagger e^{-i({\bf{kr}}-\omega t)}\right)\\
&+\left[\hbar\omega + \frac{e^2 \alpha^2}{mc^2}\right]
\left(\frac{1}{2}+a^\dagger a\right)\\
&+\frac{e^2 \alpha^2}{2 m c^2}\frac{1-\xi^2}{1+\xi^2}\left[a^2 e^{2\,i({\bf{kr}}-\omega t)} + (a^\dagger)^2 e^{-2\,i({\bf{kr}}-\omega t)}\right].
\end{aligned}
\end{eqnarray}
In order to eliminate the exponential position dependence, the following unitary transformation is used:
\bea\label{pwtrf}
U=e^{i{({\bf{kr}}-\omega t)}( a^\dagger a+\frac{1}{2})}.
\eea
Acting with the transformation $U$ on $H$ results in   
\bea\label{pwHamiltrf}
\!\!\!\!\!\!\!\! UH_{pw}U^\dagger&=&
U\frac{\bol{p}^2}{2 m}U^\dagger-
\frac{e\alpha}{m c}U\bol{p}\left(\mathbfcal{E}  a U^\dagger e^{i({\bf{kr}}-\omega t)}\right.\nn
&&+ \mathbfcal{E}^* a^\dagger U^\dagger \left.e^{-i({\bf{kr}}-\omega t)}\right)
+\left[\hbar\omega + \frac{e^2 \alpha^2}{mc^2}\right]
\left(\frac{1}{2}+a^\dagger a\right)\nn
&&+\frac{e^2 \alpha^2}{2 m c^2}\frac{1-\xi^2}{1+\xi^2}\left[U a^2 U^\dagger e^{2\,i({\bf{kr}}-\omega t)}\right.\nn\setcounter{equation}{4}
&&\left. + U (a^\dagger)^2 U^\dagger e^{-2\,i({\bf{kr}}-\omega t)}\right].
\eea
The term containing the number operator $a^\dagger a$ transforms trivially. All the other terms are considered separately in the following. In order to compute the action of $U$ the Baker-Hausdorff-Campbell identity is used, i.e.,
\bea
e^X Y e^{-X}= Y + [X,Y] + \frac{1}{2} [X,[X,Y]] + \cdots\,\, ,
\eea
where $X$ and $Y$ are operators. Applying the lemma first to the kinetic term gives
\bea
U{\bf{p}}^2U^\dagger &=&{\bf{p}}^2 + i {\bf{k}}\left( a^\dagger a+\frac{1}{2}\right) \left[{\bf{r}},{\bf{p}}^2\right]\nn \setcounter{equation}{6}
&&+ \frac{ (i {\bf{k}})^2 }{2}\left( a^\dagger a+\frac{1}{2}\right)^2 \left[{\bf{r}},\left[{\bf{r}},{\bf{p}}^2\right]\right]+\cdots.
\eea 
The commutators give $\left[{\bf{r}},{\bf{p}}^2\right]=2 i\hbar {\bf{p}}$
and $\left[{\bf{r}}\left[{\bf{r}},{\bf{p}}^2\right]\right]=(i \hbar)^2$. The higher-order terms are identically zeros since the second commutator results in a $c$-number. Thus,
\bea
U\frac{{\bf{p}}^2}{2 m}U^\dagger &=&\left[\frac{{\bf{p}}^2}{2m}- \frac{\hbar {\bf{k}}{\bf{p}}}{m} \left( a^\dagger a+\frac{1}{2}\right) + \frac{\hbar^2 k^2}{2m} \left( a^\dagger a+\frac{1}{2}\right)^2\right]\nn \setcounter{equation}{7}
&=&\frac{1}{2m}\left[{\bf{p}}-\hbar {\bf{k}} \left( a^\dagger a+\frac{1}{2}\right)\right]^2.
\eea
The term proportional to $U{\bf{p}}\mathbfcal{E}aU^\dagger$ can be rewritten as $U{\bf{p}}U^\dagger\mathbfcal{E}UaU^\dagger$; hence, the transformation for ${\bf{p}}$ and $a$ can be considered separately:
\bea
U{\bf{p}}U^\dagger= {\bf{p}} +  i {\bf{k}}\left( a^\dagger a+\frac{1}{2}\right) \left[{\bf{r}},{\bf{p}}\right]={\bf{p}} - \hbar {\bf{k}}\left( a^\dagger a+\frac{1}{2}\right).\nn\setcounter{equation}{8}
\eea
Higher-order terms vanish identically since $\left[{\bf{r}},{\bf{p}}\right]$ gives already a $c$-number. In fact the product with the polarization is
\bea
U{\bf{p}}U^\dagger\mathbfcal{E}={\bf{p}}\mathbfcal{E},
\eea 
since the wavenumber vector is orthogonal to the polarization, i.e., ${\bf{k}} \mathbfcal{E}=0$. Hence
$U{\bf{p}}\mathbfcal{E}aU^\dagger={\bf{p}}\mathbfcal{E} UaU^\dagger$.\\
The remaining terms contain $a^{(\dagger)}$ in linear and quadratic order. Their transformation is discussed in the following. The transformation of $a$ reads
\bea\label{bha}
\!\!\!\!\!\!\!\!\!\!\!\!\!\! U a U^\dagger&=& a + i({\bf{kr}}-\omega t) [a^\dagger a, a]\nn\setcounter{equation}{10}
&&+ \frac{[i({\bf{kr}}-\omega t)]^2}{2}[a^\dagger a, [a^\dagger a, a]]+\cdots.
\eea
Since $[a^\dagger a, a]=-a$, by using induction it is not hard to see that the $n$th term is $(-1)^n a$. Thus, the expression in \eqref{bha} collapses to
\bea\label{sumbha}\!\!\!\!
U a U^\dagger &=& a\left\{1+\left[-i({\bf{kr}}-\omega t)\right]+\frac{\left[-i({\bf{kr}}-\omega t)\right]^2}{2}+\cdots\right\}\nn\setcounter{equation}{11}
&=&a e^{-i({\bf{kr}}-\omega t)}.
\eea
The two exponential factors in \eqref{sumbha} and \eqref{pwHamiltrf} cancel out each other, leaving only the operator $a$ behind. Using the same procedure for $a^\dagger$ gives
\bea\label{sumbhad}\setcounter{equation}{12}
U a^\dagger U^\dagger&=& a^\dagger + i{({\bf{kr}}-\omega t)} [a^\dagger a, a^\dagger]\\
&&+ \frac{[i({\bf{kr}}-\omega t)]^2}{2}[a^\dagger a, [a^\dagger a, a^\dagger]]+\cdots=a^\dagger e^{i{({\bf{kr}}-\omega t)}}\nonumber.
\eea
The commutator in this case was $[a^\dagger a, a^\dagger]=a^\dagger$, and for the $n$th term $a^\dagger$. Again, the exponential factors in \eqref{sumbhad} and \eqref{pwHamiltrf} cancel out each other.\\
Regarding the quadratic terms in $a^{(\dagger)}$, in these cases the exponential factors of $e^{\pm 2 i({\bf{kr}}-\omega t)}$ must be eliminated:
\bea\label{bhaaadad}
\!\!\!\!\!\!\!\!U a^2 U^\dagger&=& a^2 + i{({\bf{kr}}-\omega t)} [a^\dagger a, a^2]\nn
&&+ \frac{[i({\bf{kr}}-\omega t)]^2}{2}[a^\dagger a, [a^\dagger a, a^2]]+\cdots,\nn
\!\!\!\!\!\!\!\!\!\!\!\!\!\!U (a^{\dagger})^2 U^\dagger&=& (a^{\dagger})^2  + i({\bf{kr}}-\omega t) \left[a^\dagger a, (a^{\dagger})^2 \right]\setcounter{equation}{13}\\ 
&&+ \frac{[i({\bf{kr}}-\omega t)]^2}{2}\left[a^\dagger a, \left[a^\dagger a, (a^{\dagger})^2\right]\right]+\cdots\nonumber.
\eea
And the commutators are
\bea
[a^\dagger a,a^2]&=&\left([a^\dagger,a]a  +a [a^\dagger,a]\right)a=-2a^2,\nn \setcounter{equation}{14}
\!\!\!\!\!\!\!\!\!\!\!\!\!\!\!\!\!\!\!\!\!\!\!\!\left[a^\dagger a,(a^{\dagger})^2\right]&=&a^{\dagger}\left(a^\dagger[a,a^\dagger] + [a,a^\dagger]a^\dagger\right)=2(a^\dagger)^2.
\eea 
It is not hard to see that the $n$th term gives $(\pm2)^n a^2$. Substituting these findings back to the sum gives
\bea\label{sumbhaaadad}\!\!\!\!\!\!\!\!
U a^2 U^\dagger&=&a^2\left\{1+[-2 i({\bf{kr}}-\omega t)]+\frac{[- 2 i({\bf{kr}}-\omega t)]^2}{2}+\cdots\right\}\nn
&=&a^2 e^{-2i({\bf{kr}}-\omega t)},\nn \!\!\!\!\!\!\!
U (a^{\dagger})^2 U^\dagger&=&(a^{\dagger})^2\left\{1+[2 i({\bf{kr}}-\omega t)]+\frac{[2 i({\bf{kr}}-\omega t)]^2}{2}+\cdots\right\}\nn\setcounter{equation}{15}
&=&(a^{\dagger})^2 e^{2i({\bf{kr}}-\omega t)}.
\eea
The exponential factors in \eqref{sumbhaaadad} and \eqref{pwHamiltrf}, like for the linear terms, cancel out each other.\\
Hence, collecting all the terms together, the transformed Hamiltonian reads
\begin{eqnarray}\label{pwHamiltrfje}
\begin{aligned}\setcounter{equation}{16}
UH_{pw}U^\dagger&=
\frac{1}{2m}\left[{\bf{p}}-\hbar {\bf{k}} \left( a^\dagger a+\frac{1}{2}\right)\right]^2-
\frac{e\alpha}{m c}\bol{p}\left(\mathbfcal{E}  a  + \mathbfcal{E}^* a^\dagger \right)\\
&+\left[\hbar\omega + \frac{e^2 \alpha^2}{mc^2}\right]
\left(\frac{1}{2}+a^\dagger a\right)\\
&+\frac{e^2 \alpha^2}{2 m c^2}\frac{1-\xi^2}{1+\xi^2}\left[a^2 + (a^\dagger)^2 \right].
\end{aligned}
\end{eqnarray} 
It is clear that the transformation canceled all the exponentials; however, the kinetic term of the charge has been modified, too. It is possible to rearrange the Hamiltonian in the following way ($\hat{n}=a^\dagger a$):
\bea\label{pwHamiltrfje2}\setcounter{equation}{17} \!\!\!\!
UH_{pw}U^\dagger&=&
\frac{{\bf{p}}^2}{2m}-
\frac{e\alpha}{m c}\bol{p}\left(\mathbfcal{E}  a  + \mathbfcal{E}^* a^\dagger \right)\\
&&+\left[\hbar\omega + \frac{e^2 \alpha^2}{mc^2} - \frac{\hbar {\bf{k}}{\bf{p}}}{m} + \frac{\hbar^2 k^2}{2m} \left( \hat{n}+\frac{1}{2}\right) \right]
\left(\hat n +\frac{1}{2}\right)\nn \setcounter{equation}{17}
&&+\frac{e^2 \alpha^2}{2 m c^2}\frac{1-\xi^2}{1+\xi^2}\left[a^2 + (a^\dagger)^2 \right].\nonumber
\eea 
Comparing \eqref{pwHamiltrfje2} to \eqref{dahamil}, two new terms appear in the coefficient of $(\hat{n}+1/2)$ in \eqref{pwHamiltrfje2}, namely,
\bea\label{newtermss}
\setcounter{equation}{18}
 -\frac{\hbar {\bf{k}}{\bf{p}}}{m} \qquad\text{ and } \qquad \frac{\hbar^2 k^2}{2m} \left( \hat{n}+\frac{1}{2}\right).
\eea
These two terms clearly modify the dispersion relation that was found in \eqref{Omega}. However, the detailed analysis of this modified dispersion is beyond the scope of the present paper. On the other hand, there are circumstances where \eqref{Omega} remains valid. The term $\propto {\bf{k}}{\bf{p}}$ vanishes if ${\bf{k}}$ is perpendicular to ${\bf{p}}$ or when the ${\bf{p}}\to {\bf{0}}$ limit is taken. The latter scenario describes a system where the EM mode interacts with a charge that is uniformly distributed in the volume $V$. This limit is the subject of the analysis performed in \sect{dispsec}. The remaining Hamiltonian reads 
\bea\label{pwHamiltrfje3}
\lim\limits_{{\bf{p}}\to {\bf{0}}}UH_{pw}U^\dagger&=&
\left[\hbar\omega + \frac{e^2 \alpha^2}{mc^2} + \frac{\hbar^2 k^2}{2m} \left( \hat{n}+\frac{1}{2}\right) \right]
\left(\hat n +\frac{1}{2}\right)\nn \setcounter{equation}{19}
&&+\frac{e^2 \alpha^2}{2 m c^2}\frac{1-\xi^2}{1+\xi^2}\left[a^2 + (a^\dagger)^2 \right].
\eea
The second new term is of ${\mathcal{O}}(\hbar^2)$; hence this can be considered negligible compared to the ${\mathcal{O}}(\hbar)$ terms for not very large wavenumber and photon number, for which it could dominate. Altogether the Hamiltonian is
\bea\label{pwHamiltrfje4}
\lim\limits_{{\bf{p}}\to {\bf{0}}}UH_{pw}U^\dagger&=&
\left[\hbar\omega + \frac{e^2 \alpha^2}{mc^2} + {\mathcal{O}}(\hbar^2) \right]
\left(\hat n +\frac{1}{2}\right)\nn \setcounter{equation}{20}
&&+\frac{e^2 \alpha^2}{2 m c^2}\frac{1-\xi^2}{1+\xi^2}\left[a^2 + (a^\dagger)^2 \right],
\eea
which reproduces $\lim_{{\bf{p}}\to {\bf{0}}} H$ in \eqref{dahamil} up to a negligible term of ${\mathcal{O}}(\hbar^2)$. This shows that all the conclusions drawn for the dispersion relation obtained from the Hamiltonian with dipole approximation remains valid for the plane wave vector potential, too, under the condition that ${{\bf{p}}\to {\bf{0}}}$ and for not too large wavenumber and photon number. It is interesting to note that the zero-momentum limit and the action of $U$ are not interchangeable:
\bea
U\lim\limits_{{\bf{p}}\to{\bf{0}}}H_{pw}U^\dagger&=&\lim\limits_{{\bf{p}}\to{\bf{0}}}H,\nn\setcounter{equation}{19}
\lim\limits_{{\bf{p}}\to {\bf{0}}}UH_{pw}U^\dagger&=&\lim\limits_{{\bf{p}}\to{\bf{0}}}H+{\mathcal{O}}(\hbar^2)\approx\lim\limits_{{\bf{p}}\to{\bf{0}}}H.
\setcounter{equation}{21}
\eea
In the first case the equality is exact, whereas in the second case it is only approximate.

\section{Derivation of the repulsive force between two parallel plates}\label{forceapp}
The vacuum energy for the general EP case reads as
\bea
E= \frac{1}{2}\hbar \sqrt{\omega ^2+\omega_p^2 \left(1+\frac{\xi^2}{\left(\xi^2+1\right)^2}\frac{\omega_p^2}{\omega^2}\right)},
\eea
which simplifies to the second term in $E_c(n_a=0)$ and $E_l(n_b=0)$ in \eqref{energies} when taking the limits $\xi\to1$ and $\xi\to0$, respectively.
The plasma frequency $\omega_p$ encapsulates the finite-volume dependence
\bea\label{pfr}
\omega_p= \frac{ 2 \sqrt{\pi}e}{\sqrt{m \,{\mathcal{A}}\,d} },
\eea
where $\mathcal{A}$ is the surface area of the two parallel plates and $d$ is their distance from each other. The force between the two plates is computed as
\bea\label{forcecomp}
F=-\frac{\partial E}{ \partial d}.
\eea
The only term that depends on the distance is the plasma frequency. Its derivative reads 
\bea
\frac{\partial\,\omega_p}{\partial\,d}=-\frac{ 2 \sqrt{\pi}e}{\sqrt{m \,{\mathcal{A}}} }\frac{1}{d^{3/2}}=-\frac{1}{2} \frac{\omega_p}{d}.
\eea
Thus, the force by using \eqref{forcecomp} is
\bea\label{forcew}
F&=&-\frac{\hbar \, \omega_p \frac{\partial\,\omega_p}{\partial\,d}  \left(1+\frac{2 \xi ^2}{\left(\xi ^2+1\right)^2}\frac{ \omega_p^2}{ \omega^2}\right)}{2 \sqrt{\omega^2+\omega_p ^2\left(1+\frac{\xi ^2}{\left(\xi ^2+1\right)^2}\frac{ \omega_p^2}{ \omega^2}\right)}}\nn \setcounter{equation}{5}
&=&\frac{\hbar \,\omega_p\left(1+\frac{2 \xi ^2}{\left(\xi ^2+1\right)^2}\frac{ \omega_p^2}{ \omega^2}\right)}{4 \sqrt{\frac{\omega^2}{\omega_p ^2}+\left(1+\frac{\xi ^2}{\left(\xi ^2+1\right)^2}\frac{ \omega_p^2}{ \omega^2}\right)}}\frac{1}{d}.
\eea
Here, it must be remembered that the plasma frequency depends on the geometric parameters $d$ and $\mathcal{A}$, i.e., $\omega_p=\omega_p(d,\mathcal{A})$. By substituting \eqref{pfr} into \eqref{forcew} the explicit distance and surface area dependence can be obtained. However, if the $\Delta d$ change in the distance does not modify the plasma frequency considerably, i.e., $\omega_p(d+\Delta d)\approx \omega_p$, then \eqref{forcew} gives the frequency-dependent repulsive force between the plates that scales as $\propto 1/d$ with the distance. For fixed $d$ the limit $\lim_{\omega\to \infty} F = 0$ for all $\xi$, on the other hand, the limit $\lim_{\omega\to 0} F = \infty$ for $0<\xi\leq 1$. In the case of LP ($\xi=0$), the limit $\lim_{\omega\to 0} F = \hbar\,\omega_p/4 d$, which is the maximum of the force $F$ for LP.\\
At the minimum of the zero-point energy, $\omega^*=\omega_p\sqrt{\xi/(1+\xi^2)}$, the repulsive force in \eqref{forcew} takes the form
\bea\label{dadsd}
F^*=\frac{\kappa}{4} \frac{\hbar\,\omega_p(d,{\mathcal{A}})}{d}=\frac{\kappa}{2}\sqrt{\frac{\pi R_B}{\mathcal{A}}} \frac{e^2}{d^{3/2}},
\eea
where $\kappa=(1+\xi)/\sqrt{1+\xi^2}$ and $R_B=\hbar^2/m e^2$ is the Bohr radius. For $\xi=0$ this coincides with the force in the $\omega\to 0$ limit.

\end{document}